\documentclass[11pt]{article}

\newsavebox{\PSLASH}
\sbox{\PSLASH}{$p$\hspace{-1.8mm}/}

\begin{document}
\title{Winding Number of Fractional Brownian Motion}
\author{M. A. Rajabpour\footnote{e-mail: rajabpour@ipm.ir} \\ \\
Institute for Studies in Theoretical Physics and Mathematics,
Tehran 19395-5531, Iran} \maketitle
\begin{abstract}
We find the exact winding number distribution of Riemann-Liouville
fractional Brownian motion for large times in two dimensions using
the propagator of a free particle. The distribution is similar to
the Brownian motion case and it is of Cauchy type. In addition we
find the winding number distribution of fractal time process, i.e.
time fractional Fokker-Planck equation, in the presence of finite
size winding center.
  \vspace{5mm}%
\end{abstract}

\section{Introduction}
Fractional diffusion equations are the basic methods to describe a
large class of non-equilibrium phenomena which show a power law mean
square displacement. In the normal case the mean square displacement
is asymptotically linear in time, i.e. $<x^{2}> \sim t$, but for the
anomalous cases it has a non-linear behavior. There are many
examples showing anomalous displacement e.g: protein dynamics
\cite{glocke1}, relaxation processes and reaction kinetics of
proteins \cite{glocke2}, two dimensional rotating flows
\cite{solomon}, porous glasses \cite{stapf}, intercellular transport
\cite{caspi} and so on. There are many different kinds of fractional
processes but only two of them have more application and so have
been studied in more detail: fractional Brownian motion
(FBM)\cite{M} and the fractal time process with fractional
Fokker-Planck equation \cite{ Montrol}. Both processes are
non-Markovian and have the power law mean square displacement
$<x^{2}> \sim t^{2H}$, where $H$ is a real number, but they are
fundamentally different \cite{lutz}.

The above processes can describe anomalous diffusion of a particle
in time or diffusion of a macromolecule, in other words it is
possible to look at these processes as models for dynamics of
macromolecules . We have a fractional process which describes the
diffusion of a polymer in a specific media. One of the most
important characteristics of a polymer is the winding number
distribution which is the simplest quantity describing the
entanglement of the macromolecule with point-like molecules or
molecules with finite size.

As a simple way to define winding number let's see the process as a
two dimensional random walker. A two dimensional random walker that
starts from the neighborhood of a point in the plane tend to follow
a path that wraps around that point, the measure of the wrapping is
given via the winding angle, which is the angle around the reference
point swept out by walker. Winding angles of paths are of great
interest not only from mathematical point of view but also because
of their application in physics of polymers, flux lines in high
temperature superconductors and quantum Hall effect \cite{degennes,
nelson,wilkzek}. Winding number of Brownian motion was calculated
long time a go by Spitzer \cite{spitzer}. He showed that the
distribution of winding number is of Cauchy type, for some other
proofs see \cite{edwards,prager,wiegel,kardar,K}. Such calculation
is missing for FBM as a continuous generalization of Brownian motion
\cite{M}. In this paper we will find the exact asymptotic
distribution function of winding number of two dimensional FBM in
the infinite plane. In addition we will derive the same quantity for
the fractal time process in the infinite plane with and without
finite size winding center.

The paper is organized as follows: in the next section following the
same method as in \cite{Grosberg} we will find a simple formula
similar to Spitzer's result for the winding number distribution of
FBM. In section $2$ we will calculate the same quantity for the
fractal time process, in this case we will also calculate the
winding number distribution in the presence of an obstacle which is
a good model to describe the entanglement of a polymer in the
presence of finite size molecule. Finally the last section
summarizes our results.

\section{Winding Number Distribution of FBM}
\setcounter{equation}{0}
 Fractional Brownian motion is defined as a
continuous stochastic process with zero mean value and
$<B_{H}^{2}(t)>\sim t^{2H}$, with $0< H < 1$. To calculate the
winding number of such a process we will follow the reference
\cite{Grosberg} which is based on the methods introduced in the
papers \cite{spitzer,edwards,prager,wiegel} to derive the winding
number distribution of Brownian motion. First we need to have the
probability of finding the particle in $\textbf{r}$ at the time $t$.
This probability is calculated for the Riemann-Liouville fractional
Brownian motion in \cite{sebastian,oshanin}
 defined as:
\begin{eqnarray}\label{RLFB}
B_{H}(t)=D\int_{0}^{t}\frac{d\tau \xi(\tau)}{(t-\tau)^{1/2-H}},
\end{eqnarray}
where $\xi(\tau)$ is Gaussian, delta correlated noise
$<\xi(t)\xi(t')>=\delta(t-t')$, and $D=\frac{1}{\Gamma(H+1/2)}$.
This process is not markovian and does not have stationary
increments, therefore it is different from FBM with stationary
increments which is more investigated in the mathematical
literature. The Green function of this process satisfies the
following diffusion equation
\begin{eqnarray}\label{green FBM}
\frac{\partial}{\partial t}G_{H}(\textbf{r}_{1},\textbf{r}_{2},t)-
\frac{1}{2}D^{2}
t^{2H-1}\nabla^{2}_{\textbf{r}_{1}}G_{H}(\textbf{r}_{1},\textbf{r}_{2},t)=
\delta (\textbf{r}_{2}-\textbf{r}_{1})\delta(t),
\end{eqnarray}
where $\nabla^{2}_{\textbf{r}_{1}}$ is the Laplace operator acting
on $\textbf{r}_{1}$. The above equation can be considered as the
effective differential equation describing propagator of FBM (as
already proposed in \cite{wang}\footnote{it seems that this equation
is effectively true and gives the right variance, Green's function
and the time-dependence of the survival probability \cite{oshanin}.
The survival probability is the long time asymptotics of the
probability $P_{t}$ that the FBM does not scape from a fixed
interval up to time $t$.}). Following \cite{Grosberg} one can write
the solution of the above equation as
\begin{eqnarray}\label{propagator}
G_{H}(r_{1},r_{2},\theta,t)=\frac{1}{2\pi
}\int_{0}^{\infty}\int_{0}^{\infty}\exp(-\frac{D^{2}t^{2H}\kappa^{2}}{2})\cos(\mu\theta)J_{\mu}(\kappa
r)J_{\mu}(\kappa r')\kappa d\kappa d\mu,
\end{eqnarray}
where $J_{\mu}(s)$ is the Bessel function of the first kind and
$r_{1}$ and $r_{2}$ are the moduli of $\textbf{r}_{1}$ and
$\textbf{r}_{2}$ respectively. It is possible to look at $\kappa$ as
the eigenvalue of the Laplacian then the above equation is just a
bilinear expansion over the corresponding eigenfunctions. The above
equation is similar, up to the power of $t$, to the Brownian motion
counterpart and can be checked straightly by putting it in to the
equation (\ref{green FBM}).

It is worth mentioning exact meaning of the Green function, i.e.
$G_{H}(r_{1},r_{2},\theta,t)$, it is the statistical weight of
trajectories of two dimensional FBM that start at a point
$\textbf{r}_{1}$ away from the winding center, here origin, and
arrive at another point $\textbf{r}_{2}$ after time $t$ after
winding angle $\theta$ around origin. If we drop the restriction on
the winding around origin then we can substitute the integral on
$\mu$ by the sum and after calculating the sum we will get Gaussian
distribution function for the trajectories of two dimensional FBM
that start at a point $\textbf{r}_{1}$ and arrive at point
$\textbf{r}_{2}$ at time $t$, the variance of this Gaussian function
is $t^{H}$, see \cite{sebastian,oshanin}. The reason for including
all of the positive $\mu$s in the equation (\ref{propagator}) is as
follows: we are trying to calculate winding number distribution in
our problem so technically $\theta$ is different from $\theta \pm
2\pi n$, $n$ is integer, putting sum in the equation
(\ref{propagator}) will not consider this difference. However we
should mention that it is also possible to get the true answer by
taking the sum and following Kholodenco's method for finding winding
number distribution \cite{K}. The other point to mention is, if we
include also negative $\mu$s in the integral then, since
$J_{\mu}(s)$ for small $s$ is singular for $\mu<0$, we will not get
finite propagator for $r_{1}-r_{2}\rightarrow0$. The equation
(\ref{propagator}) is also consistent with propagators in
\cite{K,sebastian,oshanin}.

To get winding number distribution we should evaluate the integral
(\ref{propagator}) for large times. The integration over $\kappa$ is
truncated at $\kappa^{2}\leq \frac{2}{D^{2}t^{2H}}$, and so we can
replace the Bessel function by the first term of its expansion, i.e.
$J_{\mu}(s)\simeq\frac{1}{\Gamma(1+\mu)}(\frac{s}{2})^{\mu}$. The
integration over $\kappa$ gives
\begin{eqnarray}\label{propagator2}
G_{H}(r_{1},r_{2},\theta,t)\simeq \frac{1}{2\pi
D^{2}t^{2H}}\int_{0}^{\infty}(\lambda_{H})^{\mu}\frac{\cos(\mu\theta)}{\Gamma(1+\mu)}d\mu,
\end{eqnarray}
where $\lambda_{H}=r_{1}r_{2}/(2D^{2}t^{2H})$. If we choose
$r_{2}=\bar{r}t^{H}+r_{1}$ then for large $t$ we have
 $\lambda_{H}\simeq \frac{r_{1}\bar{r}}{2D^{2}t^{H}}$, which goes
 to zero for large times. If we consider just small $\lambda_{H}$ in the
 integrand then the integral will be dominated by small $\mu$. The
 first order approximation gives
\begin{eqnarray}\label{distribution3}
G_{H}(r_{1},r_{2},\theta,t)\simeq \frac{1}{4\pi D
t^{2H}}\frac{\ln(\frac{1}{\lambda_{H}})}{(\ln(\frac{1}{\lambda_{H}}))^{2}+\theta^{2}}.
\end{eqnarray}

To get the distribution for $\theta$, since $\lambda_{H}$ is small
we can use $\ln \lambda_{H}\simeq -\ln t^{H}$, after renormalization
we have a Cauchy type distribution
\begin{eqnarray}\label{distribution5}
G(x=\frac{\theta}{H \ln t})=\frac{1}{\pi}\frac{1}{1+x^{2}},
\end{eqnarray}
for $t$ goes to infinity. For $H=\frac{1}{2}$ this is the same as
Spitzer's result. The same result is accessible for the other
generalization of Brownian motion defined by
$B_{H}(t)=D\int_{0}^{t}\frac{d\tau \xi(\tau)}{\tau^{1/2-H}}$, see
\cite{sebastian}, with the same constant $D$. This process has the
same propagator but has a different autocorrelation
\cite{sebastian,Cherayil}. Since we just need the propagator to find
winding number distribution, the winding distribution of this
process is the same as Riemann-Liouville fractional Brownian motion.

It seems that the same calculation should be tractable for different
boundary conditions if the propagator of FBM could be calculated, of
course translational invariance plays crucial role in the above
argument. The boundary condition corresponding to the problem of
walker in the presence of finite size winding center is equivalent
to solving the equation (\ref{green FBM}) for cylindrical boundary
condition with zero generating function on the boundary. We have
unfortunately not been able to provide a solution for this problem.
However it is possible to do the calculation for the fractal time
process, fractional Fokker-Planck equation, which is the main
subject of the next section.

\section{Winding Number Distribution of Fractional Time Process}
\setcounter{equation}{0}

To define fractional time process or fractal time random walk we
will follow the approach of \cite{Metzler}. Consider a continuous
time random walk so that the waiting time between two jumps and the
length of the jumps come from a special pdf, $\psi(x,t)$. Then one
can get the jump length pdf, $\lambda(x)$, and waiting time pdf,
$w(t)$, just by integrating of $\psi(x,t)$ on $t$ and $x$
respectively. Using $\psi(x,t)$, the fractal time random walk can be
described by the following master equation
\begin{eqnarray}\label{master}
\eta(x,t)=\int_{-\infty}^{\infty}dx'\int_{0}^{\infty}dt'\eta(x',t')\psi(x-x',t-t')+\delta(x)\delta(t),
\end{eqnarray}
where in the process we have taken into account $w(t)\sim
t^{-(1+2H})$, and the jump's length variance is also finite. We are
not going to describe all of the interesting properties of this
process, the only thing we would like to mention is that this
process has
 the following fractional Fokker-Planck equation \cite{Metzler}
\begin{eqnarray}\label{distribution6}
\frac{\partial}{\partial t}G_{H}(\textbf{r}_{1},\textbf{r}_{2},t)- D
\frac{\partial^{1-2H}}{\partial
t^{1-2H}}\nabla^{2}_{\textbf{r}_{1}}G_{H}(\textbf{r}_{1},\textbf{r}_{2},t)=
\delta (\textbf{r}_{2}-\textbf{r}_{1})\delta(t).
\end{eqnarray}
In the above formula we used the Riemann-Liouville fractional
derivative $\frac{\partial^{1-2H}f(t)}{\partial
t^{1-2H}}=\frac{1}{\Gamma(2H-1)}$
$\int_{0}^{t}\frac{f(\tau)d\tau}{(t-\tau)^{2-2H}}$,
 with $1/2<H<1$, see \cite{Saichev}. The corresponding derivative for $0<H<1/2$ can also be defined by adding ordinary derivative to the fractional one.
Using the above equation it is easy to find the winding number
distribution of this process in two dimensions in the presence of a
disc-like obstacle in the origin which imposing cylindrical boundary
condition for the above propagator. Following the same steps as in
\cite{rudnick} we can find the solution for cylindrical boundary
condition with zero generating function on the boundary. In the
Laplace transform space of $t$, i.e $p$, we have
\begin{eqnarray}\label{laplace space}
\nabla^{2}_{\textbf{r}_{1}}G_{H}(\textbf{r}_{1},\textbf{r}_{2},p)-\kappa
G_{H}(\textbf{r}_{1},\textbf{r}_{2},p)=-\alpha \delta
(\textbf{r}_{2}-\textbf{r}_{1}),
\end{eqnarray}
where $\kappa \simeq p^{2H}$, and $\alpha \simeq p^{2H-1}$. If we
now go to the Fourier space of $\theta$, i.e $\mu$, we will have
\begin{eqnarray}\label{Fourier space}
(\frac{\partial^{2}}{\partial
r_{1}^{2}}+\frac{1}{r_{1}}\frac{\partial}{\partial
r_{1}})G_{H}(r_{1},r_{2},p,\mu)-(\kappa^{2}+\frac{\mu^{2}}{r_{1}^{2}})G_{H}(r_{1},r_{2},p,\mu)=-\frac{\alpha}{r_{1}}\delta(r_{1}-r_{2}).
\end{eqnarray}
The solution for the above equation is well known, see for example
\cite{rudnick}, and has the following form
\begin{eqnarray}\label{distribution6}
G_{H}(r_{1},r_{2},p,\mu)\simeq K_{\mu}(\kappa
r_{2})(\frac{I_{\mu}(\kappa r_{1})K_{\mu}(\kappa R)-I_{\mu}(\kappa
R)K_{\mu}(\kappa r_{1})}{K_{\mu}(\kappa R)}),
\end{eqnarray}
where we take $r_{2}>r_{1}$ and $R$ is the radius of disc removed
from the plan and the functions $K_{\mu}$ and $I_{\mu}$ are the
modified Bessel functions. To get winding number distribution we
need to integrate over the radial coordinate of the final points.
Returning to the $\theta$ and $t$ space by inverse Fourier and
inverse Laplace transform we can write
\begin{eqnarray}\label{distribution6}
P(\theta,t)=\int_{-\infty}^{\infty}d \mu  e^{i \mu
\theta}C(\mu)(\frac{I_{\mu}(t^{-H} r_{1})K_{\mu}(t^{-H}
R)-I_{\mu}(t^{-H} R)K_{\mu}(t^{-H} r_{1})}{K_{\mu}(t^{-H} R)}),
\end{eqnarray}
where $C(\mu)=\int_{r_{1}}^{\infty}d r_{2} r_{2}K_{\mu}(t^{-H
}r_{2})$, for $r_{1}$ close to the border of obstacle. To get the
inverse Laplace transform we used steepest descent approximation
similar to \cite{rudnick} for large times which is like substituting
$\kappa$ with $t^{-H}$ after integrating on $p$ space\footnote{ For
more detail about the inverse Laplace transform see \cite{Grosberg}
and for the application of steepest descend method for inverse
Laplace transform see \cite{petrova}}. The above equation is quite
the same as the equation $(2.15)$ in \cite{rudnick} with just
modified $\kappa$, so by following Rudnick and Hu \cite{rudnick} one
can argue that the integrand vanishes exponentially at $\pm\infty$
for large times. Then it is possible to calculate the integral by
transforming the integration over $\mu$ in to an integral around a
closed contour in the complex plane as the standard contour
integration. The integral were calculated in \cite{rudnick} for
generic small $\kappa$, i.e. in the large time limit, and has the
following form
\begin{eqnarray}\label{distribution7}
P(x=\frac{\theta}{H \ln t})=\frac{\pi}{4 \cosh^{2}(\frac{\pi
x}{2})}.
\end{eqnarray}
We can not get the limit of zero winding center from the above
equation, to get this limit we should go back to the original
equation of $P(\theta,t)$ and take the limit $R\rightarrow 0$. It is
not difficult to see that just the function $I_{m}$ in the integral
will survive and we will get the similar equation as we found for
Riemann- Liouville fractional Brownian motion. It is possible also
to get the same result as (\ref{distribution7}) by following
\cite{Grosberg}, which is the same method but with slightly
different technicality.

\section{Discussion and Summary}

Using the Cauchy distribution that we found for both FBM and
fractional time process it is not difficult to show that
$<e^{in\theta}>\sim \frac{1}{t^{nH}}$. It is the same as the result
for FBM with autocorrelation
$<B_{H}(t+T)B_{H}(t)>=\frac{1}{2}((t+T)^{2H}+t^{2H}-T^{2H})$, which
has stationary increments, at large $n$ \cite{Nualart}. We should
emphasize that our result is not necessarily true for this kind of
FBM which has different kind of Fokker-Planck equation. However it
seems that the power law behavior of generating function of winding
angle at large times for large winding numbers is the general
property of different kinds of fractional Brownian processes.
Finally we think that it is interesting to calculate the same
winding distributions for other familiar situations specially those
proposed in \cite{Grosberg}, such as: paths with fixed endpoints
with and without obstacle in winding center and paths with glued
endpoints. These cases as argued in \cite{Grosberg} are more related
to the polymer applications.

It is interesting also to calculate the winding number distribution
by using other methods, specially path integral method, and see the
connection to fractional quantum mechanics as already done for the
ordinary Brownian motion in \cite{edwards, wiegel}. It is also
interesting to verify our formula by numerical studies.

In conclusion we calculated the exact winding number distribution of
Riemann-Liouville fractional Brownian motion for the point-like
winding center which is the first result in this context. We did
more general calculation for the fractal time process with finite
size winding center and showed that the equation is, up to a
parameter $H$, similar to the Brownian motion winding number
distribution. Moreover we showed that at least for large times and
large winding numbers the generating function of winding number is
power law in the absence of obstacle.

\textit{ Acknowledgment}:

I thank M. Kardar for valuable comments and J. Cardy for reading the
manuscript. I also would like to acknowledge the careful reading of
manuscript by S. Moghimi-Araghi and A. Ghaumzadeh. I am also
indebted to careful referees for very constructive comments.

\end{document}